\documentclass[12pt,a4paper]{article}
\usepackage{fullpage}
\usepackage{cite}
\usepackage{amsbsy}
\usepackage{amsfonts}
\usepackage{amssymb}
\usepackage{url}
\usepackage[dvips]{hyperref}
\begin{document}

\begin{flushright}
\textsf{15 December 2003}
\\
\textsf{hep-ph/0312180}
\end{flushright}

\vspace{1cm}

\begin{center}
\large
\textbf{Reply to ``{A remark on the ``Theory of neutrino oscillations''}''}
\normalsize
\\[0.5cm]
\large
Carlo Giunti
\normalsize
\\[0.5cm]
INFN, Sezione di Torino, and Dipartimento di Fisica Teorica,
\\
Universit\`a di Torino,
Via P. Giuria 1, I--10125 Torino, Italy
\\[0.5cm]
\begin{minipage}[t]{0.8\textwidth}
\begin{center}
\textbf{Abstract}
\end{center}
\begin{center}
I refute the criticisms of Ref.~\cite{hep-ph/0311241} raised in Ref.~\cite{hep-ph/0312151}.
\end{center}
\end{minipage}
\end{center}

With this short note I would like to answer the rather provocative
note of
Okun, Rotaev, Schepkin and Tsukerman (O.R.S.T.)
in Ref.~\cite{hep-ph/0312151}.

First, I would like to thank O.R.S.T. for
their attention to my recent paper in Ref.~\cite{hep-ph/0311241},
that is the object of their criticism.

The specific target of the O.R.S.T. criticism
is the use of the substitution
$t = x$
in the covariant derivation of the neutrino oscillation probability
in the plane wave approach.
Here $x$ is the distance traveled by the neutrino between production and detection
and $t$ is the elapsed time.

Unfortunately,
O.R.S.T.
did not notice or mention that in Ref.~\cite{hep-ph/0311241}
it is explained that
in order to justify the substitution
$t = x$
it is necessary to treat massive neutrinos as wave packets
\cite{Nussinov:1976uw,Kayser:1981ye,Giunti-Kim-Lee-Whendo-91,Giunti:1992sx,Kiers:1996zj,Kiers-Weiss-PRD57-98,Giunti-Kim-Coherence-98,Zralek-oscillations-98,Giunti:2000kw,Beuthe:2001rc,Dolgov:2002wy,Giunti:2002ee,Giunti:2002xg,Giunti:2003ax}.
I think that,
if one accepts the assumption
$t = x$,
the covariant derivation of the neutrino oscillation probability
in the plane wave approach
presented in Ref.~\cite{hep-ph/0311241}
does not present inconsistencies.

The physical reason why the substitution $t = x$ is correct
can be understood by noting that
the massive neutrino wave packets
overlap with the detection process
for an interval of time $[ t - \Delta t \,,\, t + \Delta t ]$,
with
\begin{equation}
t
=
\frac{x}{\overline{v}}
\simeq
x \left( 1 + \frac{\overline{m^2}}{2E^2} \right)
\,,
\qquad
\Delta t \sim \sigma_x
\,,
\label{001}
\end{equation}
where $\overline{v}$
is the average group velocity,
$\overline{m^2}$ is the average of the squared neutrino masses,
$E$ is the neutrino energy (neglecting mass contributions),
$\sigma_x$
is given by the spatial uncertainties of the production and detection processes
summed in quadrature
\cite{Giunti-Kim-Coherence-98}
(the spatial uncertainty of the production process
determines the size of the massive neutrino wave packets).
I also assumed that massive neutrinos are ultrarelativistic
and contribute coherently to the detection process in order
to observe the oscillatory terms of the flavor
transition probability,
for which the substitution $t = x$ is needed
in order to get the space-dependent phases
\begin{equation}
\phi_{kj}(x)
=
\frac{ \Delta{m}^2_{kj} x }{ 2 E }
=
2 \pi \,
\frac{x}{L^{\mathrm{osc}}_{kj}}
\,,
\label{003}
\end{equation}
where
$\Delta{m}^2_{kj} = m_k^2 - m_j^2$
is the difference between two squared neutrino masses
and
$L^{\mathrm{osc}}_{kj} = 4 \pi E / \Delta{m}^2_{kj}$
is the corresponding oscillation length.
It is clear that the correction $ x \overline{m^2} / 2E^2 $ to $t=x$
in Eq.~(\ref{001})
can be neglected,
because it gives corrections to the oscillation phases
which are of higher order in the very small ratios\footnote{
Since the neutrino masses are smaller than about one eV
(see Ref.~\cite{hep-ph/0310238})
and only neutrinos with energy larger than about 100 keV
can be detected
(see Ref.~\cite{Giunti:2002xg}),
we have
$ m_k^2 / E^2 \lesssim 10^{-10}$.
}
$ m_k^2 / E^2 $.
The corrections due to
$\Delta t \sim \sigma_x$
are also negligible,
because in all realistic experiments
$ \sigma_x \ll L^{\mathrm{osc}}_{kj} $,
otherwise oscillations could not be observed
\cite{Kayser:1981ye,Giunti-Kim-Lee-Whendo-91,Beuthe:2001rc,Giunti:2003ax}.
One can summarize these arguments by saying that
the substitution $t = x$ is correct
because
the phase of the oscillations
is practically constant over the interval of time in which the
massive neutrino wave packets overlap with the detection process
and it is given by
Eq.~(\ref{003})
plus negligible corrections of higher order in the neutrino masses.

Let me answer in detail to
the two false problems raised by O.R.S.T.:

\begin{enumerate}
\item
One is free to define
``so-called space velocities $\bar v = x/t$''
of massive neutrinos,
which are trivially identical.
However,
I think that such definition is useless.
Neutrinos are not classical objects,
for which $ v = x/t $.
It is pretty clear that
the uncertainty principle forbids any relation of such type.

Certainly,
the ``space velocities $\bar v = x/t$''
have no relevance at all for the phase of a quantum wave function.

The judgment
``But in this case the plane wave approximation is not valid.''
does not seem to follow from any logical consideration.
\item
The second remark does not apply to my paper in Ref.~\cite{hep-ph/0311241},
where the substitution
$t = x$
was adopted.
Without any logic,
O.R.S.T.
write that since the different prescription
$x = v_k t$
gives a wrong result, ``the
``theory'' of ref. \cite{hep-ph/0311241} is false.''
\end{enumerate}

Both the two false problems claimed by O.R.S.T.
stem from a confusion between phase velocity, group velocity
and other fanciful velocities
that continues to appear from time to time in the literature
(see the references in Ref.~\cite{Neutrino-Unbound}),
in spite of the discussion in Ref.~\cite{Giunti:2002ee}
and the fact that the difference between phase velocity and group velocity
and their physical relevance
are explained in several introductory books on Quantum Mechanics and Optics.

O.R.S.T.
conclude their note
claiming the validity of the so-called ``equal-energy assumption'',
which has been proved to be in general unrealistic
in Refs.~\cite{Giunti:2001kj,Giunti:2003ax}
on the basis of simple relativistic considerations.

O.R.S.T.
write that
``one has to assume that all three $E_k$ are equal''.
However, physics is not made from arbitrary assumptions
and one should justify them.
O.R.S.T.
reasoning seems to be:
since (in their opinion) the substitution $t=x$
``is fraught with at least two problems'',
the equal-energy assumption must be correct,
because it is the only one that produces the standard phase
in the plane wave approximation.
This is clearly a circular reasoning:
in order to get the standard phase one has to assume equal energies
and in order to justify the equal-energy assumption
one has to assume the correctness of the standard phase.

The physical reason why the equal-energy assumption is not realistic
is that
energy-momentum conservation in the production process
in general implies that
the energies of different massive neutrinos are different
\cite{Winter:1981kj,Giunti:2000kw},
as at least a subset of O.R.S.T. should know
\cite{Dolgov-Morozov-Okun-Shchepkin-97}.
Of course, energy and momentum are not exactly defined in
the production process,
in order to allow the coherent production
of different massive neutrinos, but it is pretty unlikely
that an uncertainty in energy and momentum could generate
the equal-energy constraint.
Indeed,
the neutrino wave packets have been calculated in the framework of
quantum field theory
in Ref.~\cite{Giunti:2002xg},
showing that in general the energies of different massive neutrinos are
different.

In conclusion,
I think that
the covariant derivation of the neutrino oscillation probability
in the plane wave approach
presented in Ref.~\cite{hep-ph/0311241}
is rather attractive,
because it is simple,
avoiding wave packet complications,
and uses a minimum of well-motivated assumptions.
In particular,
it shows clearly that
unrealistic assumptions about the values of the
energies and momenta of massive neutrinos
are not needed.
Of course,
I think that one is free to derive the neutrino oscillation probability
in the plane wave approach
through the equal-energy assumption
(as well as the equal-momentum assumption used in the original derivation of the
standard phase
\cite{Eliezer:1976ja,Fritzsch:1976rz,Bilenky:1976yj,Bilenky:1978nj}),
as long as one is aware of its limitations
and does not claim that it describes reality.

\begin{center}
\textbf{Note added after \href{http://arxiv.org/abs/hep-ph/0312280}{hep-ph/0312280}}
\end{center}

The authors of \href{http://arxiv.org/abs/hep-ph/0312280}{hep-ph/0312280}
seem to have problems with elementary logical reasoning
and elementary calculations.

\begin{description}
\item[Elementary Logic:]
$t=x$ is justified by a wave packet treatment.
Once this is accepted, wave packets are not needed
for the calculation of the phase at leading order in the neutrino mass contribution.
\item[Elementary Calculation:]
\begin{equation}
\left( E_k - E_j \right) x \frac{\overline{m^2}}{2E^2}
\sim
\frac{\Delta{m}^2_{kj}x}{2E} \, \frac{\overline{m^2}}{2E^2}
\ll
\frac{\Delta{m}^2_{kj}x}{2E}
\,.
\label{011}
\end{equation}
How is it possible that
``As emphasized in ref.\cite{hep-ph/0312151}, such corrections are of the same order as
the standard oscillation phase and as such are used from time to time in the
literature to modify the standard phase by the notorious factor of 2.''
[\href{http://arxiv.org/abs/hep-ph/0312280}{hep-ph/0312280}]?\footnote{
The factor of two mistake is not due to a correction to the time at which
interference is calculated,
but to a calculation of interference for different propagation times
of the different massive neutrinos
(see \cite{Giunti:2002ee}).}
\end{description}

Future comments will be posted at
\url{http://www.nu.to.infn.it/Neutrino_Oscillations/}.


\end{document}